# Observation of programmable spatiotemporal quantum-inspired bits

Long Chen[1], Zi Xuan Cai[1], Qing Chun Yin[1], Zhou He[1], Zhi Cai Yu[1], Xiaohan Cui[1], Tie Jun Cui[1,*], and Jian Wei You[1,*]

[1] State Key Laboratory of Millimeter Waves, Southeast University, Nanjing 210096, China.

* Email: tjcui@seu.edu.cn and jvyou@seu.edu.cn

Coherent superposition allows information to be encoded in the geometry of physical states[1-7]. Preparing individual states, however, does not determine how they can be organized throughout a wave, because the wave propagation and modulation constrain the underlying interference. Here we observe programmable spatiotemporal quantum-inspired bits in a space-time metasurface. Coherent synthesis from four discrete responses per polarization channel produces two-component states with the projective geometry of a pure quantum bit. Joint state-space synthesis retains the interference constraints of the shared waveform while allowing distinct polarization states to coexist at different resolved frequencies. Paired spatial and temporal codes organize these states into the prescribed trajectories, reconfigurable Stokes skyrmions and distinct polarizations at 48 experimentally resolved nonzero harmonic orders generated by a shared code pair. This connects the local state preparation to programmable field organization, establishing a physical basis for quantum-inspired information carriers whose encoding can exploit the geometrical paths, spatial winding and coordinated spectral states.

The ability to prepare and transform coherent superpositions gives physical states a geometry that can be used to encode information. In a pure two-level quantum system, the relative amplitudes and phase specify a point on Bloch sphere[1,2], and controlling these coordinates is central to quantum information processing[3-5]. A coherent classical polarization field admits the same projective description through its normalized Jones vector and Poincaré sphere [6,7]. Thus state preparation, overlap and reconstruction have direct classical counterparts, including for fields with nonseparable spatial and polarization structures[8-13]. Metasurfaces make this correspondence accessible through engineered scattering, with subwavelength elements controlling the amplitude, phase and polarization [14-18], distributing the information among polarization channels[19] and supporting reprogrammable responses[20,21]. Quantum-inspired metasurfaces use these responses to prepare the classical superpositions with a quantum-state representation[22-24]. The next question concerns the organization of those superpositions within an extended wave, where the information can depend on the relation between local state geometry and structure of the field as a whole.

Access to individual points on a state sphere leaves the spatial and spectral organization of the field undetermined. Temporal modulation already converts discrete scattering responses into complex harmonic amplitudes[25-27], enabling joint wavefront and spectral shaping[28-30] as well as polarization control at the carrier and sidebands[31-33]. Spatially structured fields provide a complementary form of organization through skyrmionic winding[34], in which a continuous Stokes map can acquire an integer degree under suitable boundary conditions[35-38]. Both programmable surface-wave textures[39,40] and Stokes textures controlled through meta-optics, illumination and propagation[35,38] demonstrate that such organization can be reconfigured. The unresolved requirement for a prescribed state distribution is to identify a compatible coherent field that a finite scattering alphabet can be actually generated. The propagation coherently combines the contributions from different aperture positions, while the temporal modulation ties the harmonic coefficients to a shared waveform, thus the common phase and intensity discarded by a local projective description remain essential synthesis variables. A design formulated in the state space must therefore retain the complex-field constraints that determine whether its target geometry is physically accessible. This requirement becomes decisive when the information carrier is a winding texture or a coordinated spectrum, because its identity depends on the arrangement of local superpositions throughout the field. For such a carrier, programming means finding a field that realizes the intended relations among its states, rather than selecting each local coordinate in isolation. The synthesis must reconcile the continuous geometry of the target with the discrete responses and finite temporal resources of the physical surface.

Here we propose programmable spatiotemporal quantum-inspired bits as locally defined by the classical states whose collective organization is controlled by a space-time metasurface. Paired codes in two separately addressable polarization channels coherently combine four calibrated reflection responses per channel, translating a finite scattering alphabet into programmable access to the state sphere. Successive preparations prescribe the temporal paths, whereas spatially distributed codes assemble the Stokes skyrmions with signed winding that can be changed on the same aperture. Joint optimization further distributes distinct states over 48 experimentally resolved nonzero harmonic orders generated by an unchanged code pair. Reconstruction of the complex polarization components tests these manifestations within one classical state description, retaining the interference constraints of their preparation. This formulation treats the state geometry as a design objective while preserving the coherent field that makes it observable. The resulting framework brings quantum-inspired state geometry into the design of extended electromagnetic (EM) fields and provides a route to information carriers defined by their collective spatial and spectral structures.

## Spatiotemporal programming of quantum-inspired bits

A quantum-inspired bit is a normalized superposition of two mutually coherent orthogonally polarized classical field components, represented geometrically by a point on the Poincaré sphere and organized, when extended over resolved coordinates, into a state trajectory or distribution. The progression in **Fig. 1a** begins with binary encoding, in which the field selects between two prescribed symbols. Mechanical reconfiguration expands this choice into sequential access to the polarization states within a selected observation mode and frequency. Spatiotemporal programming extends the controlled object to an entire distribution, specifying both of the local superpositions and the relations among states resolved in position, preparation time or frequency. Here the elementary information-bearing object is the local superposition, while spatial and spectral labels identify where that state is read out; the accessible collection is therefore determined jointly by the prescribed preparation and observation conditions.

For a populated harmonic order $m$, the transverse field at observation position $\mathbf{r}$ defines $|\psi_m(\mathbf{r};\tau)\rangle=\frac{E_{x,m}(\mathbf{r};\tau)|x\rangle+E_{y,m}(\mathbf{r};\tau)|y\rangle}{\sqrt{|E_{x,m}(\mathbf{r};\tau)|^2+|E_{y,m}(\mathbf{r};\tau)|^2}}$. Here $|x\rangle$ and $|y\rangle$ span the selected orthogonal polarization basis, and the state exists wherever the summed component intensity is nonzero. After the normalization and identification of vectors differing only by a common phase, the relative amplitude and phase remain as the two coordinates of a projective state, equivalently represented by a unit Stokes vector. This correspondence with the pure two-level quantum-state geometry motivates the quantum-inspired description, while the amplitudes themselves remain those of a directly reconstructed classical EM field (**Supplementary Notes 1** and **2**).

The space-time coding metasurface converts its discrete scattering responses into these effective superpositions through coherent synthesis (**Fig. 1b**). Two separately addressable polarization branches, each with four calibrated settings, are driven by paired codes $C_x(\mathbf{R},t)$ and $C_y(\mathbf{R},t)$ that specify the responses at aperture position $\mathbf{R}$ throughout a modulation period. Each harmonic accumulates the complex amplitudes over the sequence, with the occupation of the settings determining the carrier responses and their temporal ordering additionally controlling the sidebands. Coherent propagation within the two polarization branches then combines the contributions from the aperture to form the observed Jones field. The programmed state is consequently an emergent response of the coded aperture, whose relative component amplitudes need not coincide with any instantaneous reflection setting. This synthesis retains a physical freedom that the local state coordinates suppress, because aperture fields with the same polarization but different spatial profiles of common phase can evolve into different polarization textures after the propagation. We therefore construct and propagate both complex component fields before assigning the normalized quantum-inspired bit at each resolved observation coordinate. The calibrated response alphabet and sequence length determine which points can be sampled on the continuous state sphere, separating the geometry of the encoded states from the finite set of available control settings. Because the target fixes a component ratio rather than an absolute field, multiple complex fields can represent the same local state, and the codes select among the representatives compatible with the scattering responses and propagation.

Two temporal coordinates distinguish the synthesis, with $t$ resolving the modulation in a period and $\tau$ indexing the successive complete code preparations. Following $\tau$ at one position and frequency can produce the prescribed temporal path in **Fig. 1c**. The spatial section of the field instead exposes a map from the observation position to state sphere, allowing the programmed superpositions to acquire a collective winding (**Fig. 1d**). Periodic modulation supplies a third section through the frequencies $f_m$=$f_c$+$mf_s$, where $f_c$ and $f_s$ are the carrier and modulation frequencies, so that an unchanged code pair generates a spectrally resolved distribution of states linked by its common Fourier structure (**Fig. 1e**). Reconstructing the same two complex components in each section connects these observations to a single state geometry and makes its temporal, spatial and spectral organization experimentally accessible.

## Temporal interference expands the accessible state space

The temporal coding enlarges the set of accessible superpositions by coherently combining the four calibrated complex reflection responses of each separately addressed polarization branch over a complete modulation period. The paired sequences determine the relative amplitudes of a selected spectral component, and their successive reconfiguration traces a prescribed path through the state sphere. The trajectory is therefore assembled from the states synthesized by interference within each period, with the preparation index $\tau$ specifying their ordered progression. For the equal-slot sequence in **Fig. 2a**, the modulation period $T$=1/$f_s$ is divided into $L$ intervals with complex reflection responses $r_{p,n}$ in polarization branch $p$. Within the reduced channel model, coherent projection onto harmonic order $m$ gives

$$r_p^{(m)}=\frac{\kappa_m}{L}\sum_{n=0}^{L\text{-}1} r_{p,n}\exp(\text{-}\frac{i2\pi mn}{L}), \qquad p\in\{x,y\}. \tag{1}$$

The factor $\kappa_m$=sinc($m/L$)exp(-$i\pi m/L$) is Fourier transform of a finite slot, with sinc($u$)=sin($\pi u$)/($\pi u$). Multiplying by the fixed incident polarization amplitudes gives the reflected components at $f_c$+$mf_s$, for which a nonzero common factor $\kappa_m$ changes the spectral weight and phase while leaving the normalized state invariant. Using calibrated complex responses, the amplitude variations and phase offsets of the physical scatterers can be retained in the synthesis (**Supplementary Note 4**). At the carrier, **Eq. 1** becomes a weighted phasor sum whose occupation fractions control both constructive and destructive interference. The resulting complex amplitudes occupy the convex hull of the four calibrated responses, sampled at the rational occupation fractions allowed by a finite sequence. Pairing the two branch averages specifies the output polarization, and we select the codes by minimizing its projective mismatch with the target state (**Methods**). Longer sequences refine this sampling, whereas the response dictionary fixes the underlying accessible region (**Supplementary Note 5**). Because cancellation acts on the complex amplitudes, a sequence can suppress a polarization component while leaving the other finite even when the instantaneous reflection magnitudes are comparable, converting the discrete phase responses into effective relative-amplitude control. In the ideal quadrature alphabet, equal occupation of opposite responses cancels their carrier contribution, whereas unequal occupation leaves a controlled residual amplitude. For a fixed incident polarization, this cancellation illustrates how the discrete phase settings alter the component ratio without requiring continuously variable instantaneous reflection amplitudes.

The meridional path in **Fig. 2b** connects the opposite circular polarizations through intermediate ellipticities, with the code pairs and reconstructed ellipses identifying the preparation associated with each marked state (**Fig. 2c,d**). Along the equatorial path, the linear-polarization azimuth is rotated while the ellipticity remains zero (**Fig. 2e-g**). These paths probe the two angular coordinates of the sphere, whose latitude and longitude are twice the ellipticity angle and polarization azimuth, respectively. Theoretical, simulated and measured states are reconstructed in the same Stokes convention, allowing their geometrical agreement to be assessed directly (**Supplementary Note 6**). Each point represents a complete preparation indexed by $\tau$, thus the observed path characterizes the programmed state access without assigning an unmeasured dynamics to the switching transient. At a nonzero harmonic, the Fourier phase in **Eq. 1** distinguishes the sequences with identical occupation fractions but different slot orderings. A delay of $q$ slots in one branch multiplies its coefficient by $\exp(-i2\pi mq/L)$, changing the relative phase at order $m$ while preserving both branch magnitudes and their carrier averages. A common delay of both sequences instead contributes the same phase to the two components within each harmonic. The relative temporal displacement thus acts directly on the internal state, with a discrete and order-dependent phase increment linking a code operation to motion on the sphere. **Figure 2h-j** tests this spectral extension through separately optimized preparations at $f_c$, $f_c+f_s$ and $f_c-2f_s$, whose simulated and measured coordinates follow the corresponding targets (**Supplementary Note 14**). These measurements establish the local preparation mechanism subsequently used to organize the states into spatial textures and collective spectral distributions.

## Programmable winding of spatial state fields

Spatial programming converts the local geometry of a quantum-inspired bit into a field whose relative component amplitudes and phases map the observation plane onto the Poincaré sphere. In **Fig. 3a**, a vortex mode and an orthogonally polarized non-vortex mode supply the complementary phase and amplitude profiles needed to construct this map. The complex amplitudes at every observation point are resulted from the coherent propagation across the aperture, making the texture a property of the synthesized field as a whole. Reprogramming the same metasurface changes this mapping and its winding through the imposed scattering pattern. Skyrmionic texture combines the azimuthal phase winding with a radial redistribution of intensity between the polarization components. For an ideal rotationally symmetric configuration, the normalized state takes the form

$$|\psi(r,\varphi)\rangle=\cos\frac{\Theta(r)}{2}|x\rangle+e^{-i[\nu\varphi+\gamma]}\sin\frac{\Theta(r)}{2}|y\rangle. \quad (2)$$

The radial angle $\Theta(r)$ determines the relative intensities, while $\nu$ and $\gamma$ specify the azimuthal winding and a uniform relative-phase offset. With $S_3$=2Im($E_x E_y^*$), the corresponding Stokes field is written as **s** = (cos$\Theta$, sin$\Theta$cos[$\nu\varphi+\gamma$], sin$\Theta$sin[$\nu\varphi+\gamma$]). The azimuthal variation sweeps a circle at each radius, and the radial change in $\Theta$ carries these circles over the sphere to generate a two-dimensional state texture (**Supplementary Note 9**). When the normalized field is continuous and approaches a common state on the boundary, the observation domain can be compactified to a sphere and the map has degree

$$N_{\mathrm{sk}}=\frac{1}{4\pi}\int_D \mathbf{s}\cdot(\partial_X\mathbf{s}\times\partial_Y\mathbf{s})\, dX\, dY. \quad (3)$$

For **Eq. 2**, the degree reduces to $N_{\mathrm{sk}}=\nu[\cos\Theta(0)-\cos\Theta(R_B)]/2$, with the radial boundary $R_B$. Opposite endpoints $\Theta(0)=0$ and $\Theta(R_B)=\pi$ give $N_{\mathrm{sk}}=\nu$, whereas reversing the endpoints changes its sign and varying $\gamma$ rotates the Stokes field about its first axis without changing the degree. The topological charge consequently depends on the radial orientation and the adopted coordinate convention, and remains invariant under smooth deformations that preserve a defined map and its common boundary. Its density measures the oriented solid angle swept out by the local Stokes vector, relating the global winding directly to the projective geometry used for temporal state preparation. Different degrees therefore distinguish spatial organizations that share the same local two-component state space. For component phases $\exp(i\mathrm{l}_x\varphi)$ and $\exp(i\mathrm{l}_y\varphi)$, this convention gives the relative winding $\nu=\mathrm{l}_x-\mathrm{l}_y$. In the ideal description, different degrees cannot be connected by a continuous deformation that preserves both a defined state map and a common boundary state. Changing the degree requires these conditions to fail somewhere along such an interpolation. Reprogramming therefore selects different topological organizations, while the present measurements characterize the prepared configurations rather than the transient by which one configuration is replaced by another.

The six spatial code pairs in **Fig. 3b** target $N_{\mathrm{sk}}$=-3,-2,-1,+1,+2,+3, yielding the reconstructed integrals of -2.90,-1.95,-0.97,+0.98,+1.96,+2.91, respectively (**Fig. 3c**). The programmed field thus changes the orientation and multiplicity of its coverage of the state sphere. These finite-domain estimates can approach the integer targets, with their interpretation determined by spatial sampling, boundary states and the treatment of weak-field regions (**Methods**). The experiment demonstrates that the topology of a state distribution can be reconfigured through the same discrete responses used to prepare its local superpositions. A complementary modal description resolves how the spatial profiles occupy the two polarization branches, whose nonproportional fields cannot be factorized into a common spatial mode and a fixed Jones vector. We characterize this classical nonseparability within the selected spatial-mode and polarization subspace. The normalized modal coherency matrices in **Fig. 3d** and **3e** reconstruct the populations and relative coherence from the complex projection coefficients. The four target superpositions are prepared separately for each configuration and assessed by conditional modal fidelities (**Fig. 3f** and **Supplementary Note 10**). These projections retain relative spatial-phase information beyond the local polarization map, whereas **Eq. 3** resolves the winding of that map itself. Together, the measurements connect reconfigurable field topology to the coherent arrangement of quantum-inspired states on a common programmable aperture.

## Collective syntheses of spectrally distributed states

A single periodic preparation can distribute the quantum-inspired bits among different frequencies, with the two polarization amplitudes defining a distinct state at each populated Fourier order (**Fig. 4a**). These states form a collective synthesis problem because every harmonic derives from the same waveform and a change to one slot generally redistributes several Fourier coefficients. Extending the single-state preparations in **Fig. 2** to a prescribed spectral distribution therefore requires joint control of compatible superpositions under a common drive. We express this requirement as a weighted state mismatch for one candidate code pair C=$(C_x,C_y)$ over the selected populated orders M

$$\mathrm{L(C)}=\frac{\sum_{m\in \mathrm{M}} w_m\,[1\text{-}F_m(\mathrm{C})]}{\sum_{m\in \mathrm{M}} w_m},\qquad w_m>0. \tag{4}$$

Here $F_m=|\langle\psi_m^{\mathrm{tar}}|\psi_m(\mathrm{C})\rangle|^2=(1+\mathbf{s}_m^{\mathrm{tar}}\cdot\mathbf{s}_m)/2$ is a squared-overlap fidelity. Defined from the reconstructed classical fields, this fidelity depends only on their projective coordinates and is invariant under a common nonzero complex scale of the two polarization components. Every order is evaluated from the same candidate sequence through **Eq. 1**, with the full field propagated before normalization, so the optimization searches the spectral distributions permitted by the shared waveform. A target state fixes the component ratio while leaving their common amplitude and phase available to satisfy the field constraints, giving the state syntheses more freedom than prescribing every complex harmonic coefficient. The spectral weights remain separate observables, allowing the geometry of synthesized distribution to be assessed independently of how strongly each order is populated. The common code pair in **Fig. 4b** and **4c** generates a broad distribution over more than 200 harmonic orders in numerical synthesis, with theoretical and simulated coordinates referring to the same preparation. The accessible distributions remain constrained by the calibrated alphabet and code length, which in the held-slot boundary model impose exact relations between the Fourier orders separated by $L$ (**Supplementary Note 8**). The larger examples in **Supplementary Note 15** extend the spectral address set through longer sequences, while preserving the distinction between the number of resolved orders and the number of independently adjustable state coordinates. A relative delay between the polarization codes rotates each harmonic state by an angle proportional to its order, whereas a common delay preserves every harmonic polarization. The rotation is about the first Stokes axis, fixed by the linear-polarization basis. The same code operation therefore transforms the distribution collectively through an order-dependent geometrical action. These relations concern the separately resolved states; the common phases at different orders still determine the time-dependent field produced when the harmonics are recombined.

This extension has a direct temporal cost, because increasing $L$ at fixed slot duration increases the period and reduces the harmonic spacing. Holding the period fixed instead requires the faster switching, hence the size of the spectral address set must be considered together with its frequency spacing and usable bandwidth. Frequency resolution separates the populated orders even when their polarization states are nonorthogonal, while the ratio of the two complex harmonic amplitudes specifies the internal state at each spectral address. Spectral organization therefore combines a set of distinguishable frequency components with a programmable geometry of superpositions within those components. Experimentally, a shared code pair generates 48 resolved nonzero orders from $m$=-24 to -1 and $m$=+1 to +24, whose theoretical, simulated and measured coordinates are compared in **Fig. 4d** and **e**. The ellipses in **Fig. 4f** resolve the azimuth and axial ratio of each normalized polarization, with its signed circular component obtained from the relative complex phase. The comparison tests differential amplitude and phase across the two polarization branches, because these quantities move the state on the sphere whereas a common complex scale changes its spectral weight without changing its coordinates. Together with the temporal trajectories and spatial winding, the shared spectral preparation establishes a common route from finite scattering alphabet to programmable distributions of the quantum-inspired states.

## Conclusion

Programmable spatiotemporal quantum-inspired bits make the organization of coherent superposition a controllable property of an extended EM field. The same discrete scattering resources determine the local state coordinates and the spatial or spectral relations through which those states form a structured information carrier. The spatial winding expresses this connection geometrically, as a continuous distribution of local polarization states acquires a global topological degree under the appropriate boundary conditions. For quantum-inspired physics, the correspondence supplies an experimentally accessible setting in which the preparation errors, state overlap and geometrical structure can be examined through classical complex-field measurements. Combined with calibrated transformations of orthogonal modes, it could support the studies of state-control protocols and channel mappings that admit the same linear description in the classical and quantum systems. Additional coherent modes would extend the accessible state space, with the scope of the correspondence determined by the mode transformations and observables that the classical field can represent. Controlled averaging over preparations would additionally distinguish the geometry of a resolved coherent state from that of a statistical ensemble, connecting the programmable wave synthesis to classical representations of mixed-state evolution.

For wireless communications, the results suggest novel encoding schemes, in which a message specifies the relations among states, such as a geometrical path, a spatial winding or a coordinated spectral pattern. Their development calls for identifying which of the features remain distinguishable after channel propagation and how they can be reconstructed by a receiver. A communication alphabet built from such distributions would make the geometry of the complete received field part of the decoding problem, motivating new studies of how the channel transformations preserve or deform the relations among states. By making local state geometry and extended wave structure accessible within a programmable setting, the present work provides a physical basis for exploring information architectures organized by the coherent superposition.

# Methods

### Meta-atom architecture and EM characterization

The reflective metasurface consisted of 32×32 individually addressable meta-atoms on a 30 mm lattice, forming a 960×960 mm² aperture with two separately controlled linear-polarization branches. Each cell contained four surface-mounted SMP1345-040LF PIN diodes, with one diode pair predominantly controlling the $x$-polarized response and the other the $y$-polarized response. The four bias combinations of each pair defined a two-bit logical alphabet of calibrated reflection settings. The complete instantaneous configuration is specified by $[x_1 x_2 y_1 y_2]$, giving 16 joint bias configurations for the two branches. The multilayer printed-circuit-board structure has combined the patterned radio-frequency conductors, F4B dielectric layers, a continuous ground plane and a separate layer for direct-current bias routing. The F4B substrates had relative permittivity 2.65, loss tangent 0.001 and layer thicknesses of 0.254, 4 and 0.508 mm. Metallized vias connected the diodes to the bias network, whose routing was separated from the radio-frequency pattern.

The EM characterization is performed by CST Microwave Studio 2022. The unit-cell simulation has imposed periodic boundaries in the transverse directions and open boundaries normal to the metasurface, with separate $x$- and $y$-polarized excitations. Manufacturer-provided S-parameter files represented the forward- and reverse-biased diodes, retaining their frequency-dependent response in the EM model. In the reduced branch model, the four calibrated complex reflection coefficients form the dictionary

$$\mathrm{D}_p = \{d_{p,0}, d_{p,1}, d_{p,2}, d_{p,3}\}, \qquad p \in \{x, y\}. \tag{5}$$

Here $d_{p,j}$ denotes a static setting indexed by $j$, while $r_{p,n} = d_{p,j_{p,n}}$ is the response assigned to slot $n$ in **Eq. 1**. Both the magnitude and phase of each calibrated coefficient were retained when synthesizing the harmonic response. The compensated model instead used the 2×2 Jones response of each of the 16 joint configurations, including co-polarized terms, residual cross-polarization and dependence on both branch settings. **Supplementary Notes 3** and **12** specify the response operators and the relation between this joint description and the reduced branch model. Finite-aperture simulations used open boundaries and reproduced the applied coding patterns and illumination of the corresponding measurements. The two complex polarization components were evaluated at the experimental observation planes and frequencies, allowing finite-aperture truncation, edge scattering and inter-element interactions to enter the calculated fields.

For calculations based on the local response dictionary, the selected responses first define the reflected complex aperture field, which is then propagated to the observation coordinate before normalization. This construction retains the common spatial phase and amplitude that are absent from the local state representation. A dictionary obtained from periodic unit cells does not, by itself, include the change in local scattering caused by an arbitrary nonuniform neighbouring configuration.

## Fabrication and spatiotemporal code delivery

The metasurface was fabricated by multilayer printed-circuit-board processing. The metallic patterns and bias-routing structures were formed on the F4B substrates, followed by drilling, via metallization and surface mounting of the four diodes in each cell. The layers were aligned and assembled to maintain the designed scattering geometry and connections between the radio-frequency and bias networks. A digital control system generated the four-bit commands and delivered them through multichannel driving circuits that supplied the diode bias states. The first and second diode pairs were addressed through the $x$- and $y$-channel code streams on the common aperture.

Spatial patterns and temporal sequences were computed before acquisition and loaded into the control system. For static preparations, the prescribed four-bit matrix was held throughout measurement. For temporal and harmonic synthesis, the slot sequence was repeated with period $T$=1/$f_s$, and a new complete code pair defined each successive preparation indexed by $\tau$. Before EM measurements, the direct-current network was inspected for electrical continuity and correct diode switching in representative cells. Selected uniform configurations were measured and compared with the simulated scattering responses to verify the correspondence between the digital commands and the implemented bias settings. **Eq. 1** uses a held-slot approximation in which each selected response is constant within its assigned interval. The common factor $\kappa_m$ follows from integrating that interval and is distinct from the discrete Fourier sum over the logical settings. Finite switching transients, synchronization errors and frequency dependence beyond the calibrated range can modify the predicted harmonic amplitudes. The preparation coordinate $\tau$ labels the code applied during a measurement and does not specify the unresolved transient between two preparations.

## State objectives and joint code optimization

Candidate code pairs were constructed from the calibrated settings and selected by minimizing the state-matching objective in **Eq. 4**. The harmonic complex fields were evaluated before their normalization into Jones states and unit Stokes vectors at nonzero total intensity. For a single target, the loss is 1-$F$, ranging from zero for identical projective states to unity for orthogonal states at antipodal sphere coordinates. Its invariance under a common nonzero complex scale separates polarization-state accuracy from the absolute field magnitude. For a target set M, all harmonic coefficients were recomputed from the same candidate code pair and their state mismatches combined with the weights in **Eq. 4**, taken equal unless specified otherwise. Each target and selected order in **Fig. 2** used a separately optimized code pair. Each spectral distribution in **Fig. 4** used one shared pair optimized against the complete selected target set. **Supplementary Note 7** relates this objective to projective distance and analyzer contrasts, while **Supplementary Note 8** gives the Fourier constraints on jointly synthesized orders.

The state objective constrains the ratio of the two output amplitudes at each target, leaving their common complex scale to the field synthesis. It is undefined if both output components vanish and does not independently impose a minimum signal, conversion efficiency or leakage penalty. Any such constraints must be distinguished from the state loss. For a spatially structured preparation, the comparison refers to the propagated field at the observation plane, rather than to the local reflection coefficient of an individual meta-atom.

### Complex-field reconstruction and polarization geometry

The complex $x$- and $y$-polarized components were reconstructed with a common amplitude and phase reference. Polarization observables were calculated from the component fields at the same observation coordinate and resolved frequency. **Supplementary Note 13** analyses how the relative-channel calibration, spectral resolution and averaging enter these observables. With the positive temporal phasor convention $\exp(i\omega t)$ used in **Supplementary Note 9**, the Stokes quantities were

$$\begin{aligned} S_0 &= |E_x|^2+|E_y|^2, \quad & S_1 &= |E_x|^2-|E_y|^2, \\ S_2 &= 2\mathrm{Re}(E_x E_y^*), \quad & S_3 &= 2\mathrm{Im}(E_x E_y^*). \end{aligned} \tag{6}$$

For $S_0>0$, the normalized Stokes vector was

$$\mathbf{s}=(S_1,S_2,S_3)/S_0. \tag{7}$$

The theoretical, simulated and measured fields were expressed in the same transverse coordinate orientation and Stokes convention. The angular separation of two unit Stokes vectors was calculated as

$$\Delta\Omega=\arccos(\mathbf{s}_1\cdot\mathbf{s}_2). \tag{8}$$

The dot product was clipped numerically to [-1,1] before taking the inverse cosine to suppress floating-point round-off. For the resolved coherent states, fidelity followed the squared-overlap relation accompanying **Eq. 4**, equivalently $F=\cos^2(\Delta\Omega/2)$. Both quantities compare normalized polarization geometry without determining the absolute intensity of the reconstructed fields.

A nonzero Jones vector gives $|\mathbf{s}|=1$ algebraically, including when its reconstructed components contain deterministic measurement errors. A statistical mixture requires averaging the coherency products, rather than first averaging the complex fields and then constructing a single outer product. These operations generally yield different polarization descriptions. The handedness assigned to circular polarization additionally depends on the observation direction and must be interpreted with the stated phasor and coordinate conventions.

### Spatial state reconstruction and topological charge

Spatial preparations used paired two-dimensional codes for the two polarization branches. The resulting complex fields were sampled on a common transverse grid and converted into local normalized Stokes vectors where the field intensity was nonzero. Near field zeros, normalization amplifies errors in the reconstructed components, and the state map may become undefined. The comparisons between theory, simulation and measurement used the same state convention and spatial evaluation window. The numerical approximation to **Eq. 3** used central finite differences on the sampled Cartesian grid

$$N_{\mathrm{sk}}^{(h)}=\frac{\Delta X\Delta Y}{4\pi}\sum_{(i,j)\in D_{\mathrm{int}}}\mathbf{s}_{ij}\cdot[\frac{\mathbf{s}_{i+1,j}-\mathbf{s}_{i-1,j}}{2\Delta X}\times\frac{\mathbf{s}_{i,j+1}-\mathbf{s}_{i,j-1}}{2\Delta Y}]. \tag{9}$$

Here $\Delta X$ and $\Delta Y$ are the sampling intervals, and $D_{\mathrm{int}}$ contains interior points for which the central vector and both derivative stencils are defined. A common spatial window, transverse orientation and Stokes convention were used across the configurations, fixing the sign convention of $N_{\mathrm{sk}}$. The reported values are finite-domain estimates whose deviations from integer targets can reflect discretization, aperture truncation and nonuniform boundary states. An integer degree requires a continuously defined state map with a common boundary state. An intensity mask can create internal boundaries, and its effect cannot be represented solely by deleting terms from the sum without examining the remaining domain. The integral over a finite measured window is therefore reported separately from its target integer. A change in the azimuthal phase sign or in the orientation of the spatial axes must be propagated consistently through the reconstructed Stokes field and the integration.

**Modal coherence and conditional state fidelity**

The spatial and polarization structure was also characterized by projecting the reconstructed complex fields onto selected spatial modes over a collection domain $D$. For basis functions orthonormal under the collection inner product, the coefficient of mode $\mu$ in polarization $p$ is

$$a_{\mu p}=\int_D u_\mu^*(X,Y)E_p(X,Y)\,dX\,dY. \tag{10}$$

For the sampled templates that are not orthonormal, **Supplementary Note 10** gives a consistent construction using the Gram matrix. With square roots of the quadrature weights included in the template matrix $U$ and sampled field vectors, $G=U^\dagger U$ and $V=UG^{-1/2}$ define an orthonormal basis, and the projected coefficients are $\mathbf{a}_p=V^\dagger\mathbf{E}_p$. This construction requires linearly independent templates. Experimental and target coefficients must be expressed in the same resulting basis before matrix reconstruction or fidelity evaluation. For the selected mode pair, the ordered product basis and coefficient vector were

$$\begin{aligned}\mathrm{B}_\mathrm{l}&=(|u_0,x\rangle,|u_\mathrm{l},x\rangle,|u_0,y\rangle,|u_\mathrm{l},y\rangle),\\ \mathbf{c}_\mathrm{l}&=(a_{0x},a_{\mathrm{l}x},a_{0y},a_{\mathrm{l}y})^T.\end{aligned} \tag{11}$$

The normalized modal coherency matrix within this subspace was constructed as

$$\rho_{\mathrm{sub}}=\frac{\mathbf{c}_\mathrm{l}\mathbf{c}_\mathrm{l}^\dagger}{\mathbf{c}_\mathrm{l}^\dagger\mathbf{c}_\mathrm{l}},\qquad \mathbf{c}_\mathrm{l}^\dagger\mathbf{c}_\mathrm{l}>0. \tag{12}$$

This outer product has unit trace and rank one for every nonzero coefficient vector, including a vector affected by deterministic reconstruction errors. Its diagonal entries give normalized modal populations and its off-diagonal entries retain their relative coherences. For a normalized target in the same orthonormal basis, the conditional modal fidelity was

$$F_{\mathrm{sub}}=\langle\Psi_{\mathrm{tar}}|\rho_{\mathrm{sub}}|\Psi_{\mathrm{tar}}\rangle. \tag{13}$$

The four target superpositions in **Fig. 3f** were prepared and evaluated separately for every spatial configuration. Subspace normalization excludes field components outside the selected modal span from the fidelity comparison. If $W_{\mathrm{sub}}$ is the fraction of the reconstructed squared field norm over $D$ captured in the selected orthonormal subspace, the squared normalized overlap, over the same collection domain, with a target wholly contained in that subspace is $F_{\mathrm{full}}=W_{\mathrm{sub}}F_{\mathrm{sub}}$. The conditional value therefore describes agreement within the measured modal representation. Spatial-polarization nonseparability concerns the factorization of one classical field into mode and polarization factors, as specified in **Supplementary Note 10**.

### Harmonic resolution and spectral field weights

Periodic modulation generates integer Fourier orders, whose observable range is determined by slot duration, switching bandwidth, device dispersion, conversion into the selected orders and measurement sensitivity. For spectral synthesis, target state coordinates were assigned to the selected set M and optimized jointly for one code pair. The state loss and spectral field weights were evaluated separately, because the normalized geometry does not determine the amplitude carried by each order. At a specified observation coordinate, the relative squared-field weight was defined by

$$\widetilde{W}_m=\frac{W_m}{\sum_{k\in \mathrm{M}} W_k},\qquad W_m=|E_{x,m}|^2+|E_{y,m}|^2. \tag{14}$$

The normalization refers only to the selected harmonic set and gives a power fraction only for consistently power-normalized collection modes and amplitudes. A resolved polarization state additionally requires sufficient signal relative to measurement noise, and neither its normalized fidelity nor its relative spectral weight alone determines total conversion efficiency. The experiment resolved 48 nonzero orders from $m$=-24 to -1 and from $m$=+1 to +24 under an unchanged code pair. At each order, the calibrated two-component complex response defined the corresponding normalized quantum-inspired state. Frequency resolution and the measurement window determine whether neighbouring harmonics are separated in the reconstructed response. An unresolved combination of orders generally has a time-dependent polarization, while averaging their coherency products produces an intensity-weighted polarization mixture. The harmonic states reported here refer to individually resolved orders, so their normalization does not replace a measurement of the recombined waveform or its absolute power.

## Data availability

The data that support the findings of this study are available from the corresponding authors upon reasonable request.

## Code availability

The code that supports the findings of this study are available from the corresponding authors upon reasonable request.

## Acknowledgements

This work was supported by the National Natural Science Foundation of China (62288101 and 92167202), the National Key Research and Development Program of China (2023YFB3813100, 2022YFA1404903), the SEU Innovation Capability Enhancement Plan for Doctoral Students (CXJH_SEU 26043), Independent Research Fund of the State Key Laboratory of Millimeter Waves (Z202602-01, K201924), and the 111 Project (111-2-05).

## Author contributions

L.C., J.Y. and T.C. conceived the idea. L.C., Z.C., Q.Y., Z.H. and Z.Y. designed the experiments. L.C., X.C., J.Y. and T.C. supervised the project. L.C., Z.C., Q.Y. and Z.H. conducted the experiments, collected and analyzed the data. L.C. and Z.C. carried out the simulations, theoretical analyses, and wrote all the code. L.C., J.Y. and T.C. wrote the manuscript, with contributions from all authors.

## Competing interests

The authors declare no competing interests.

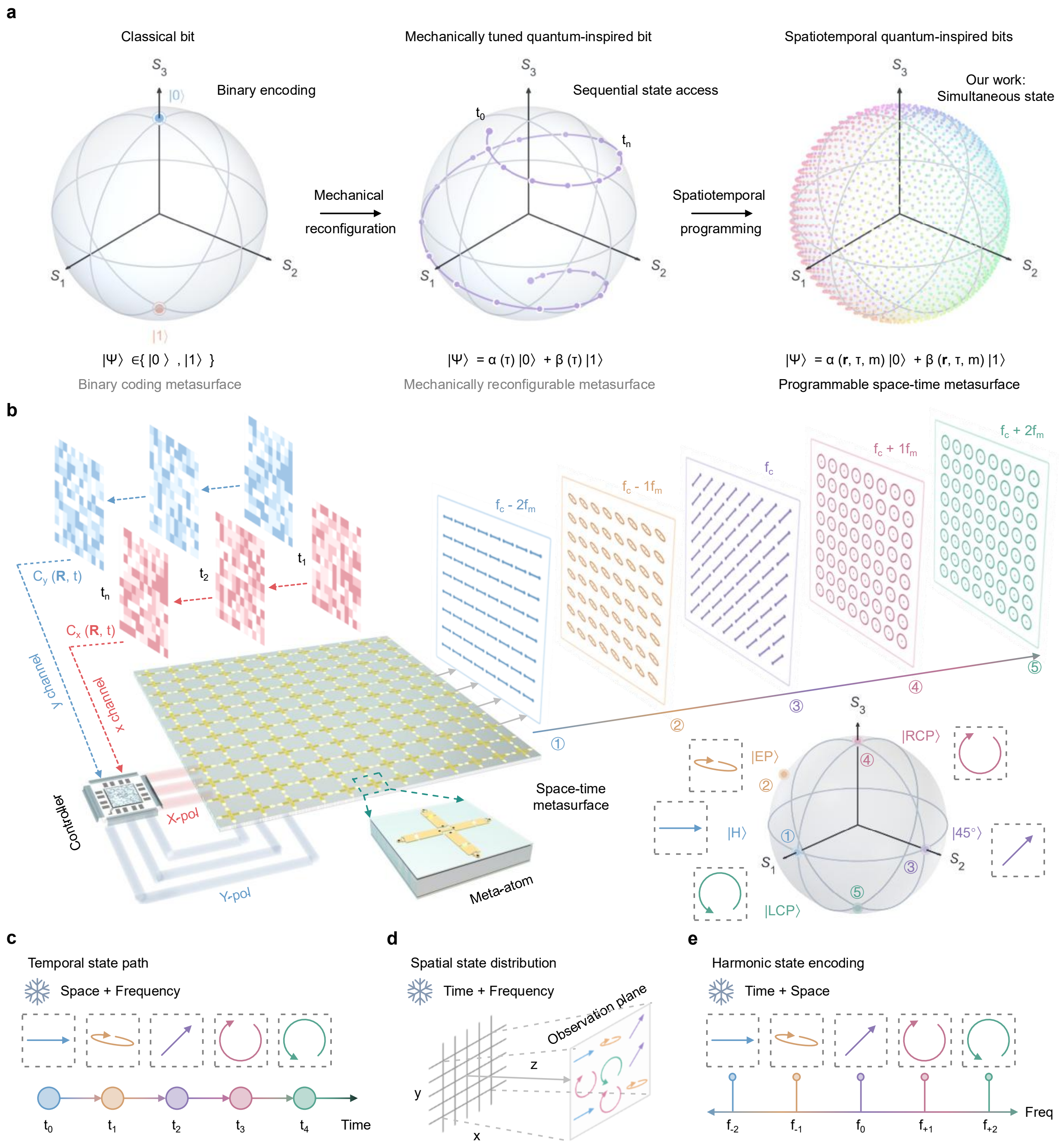

**Fig. 1 | Programmable distributions of the spatiotemporal quantum-inspired states.** **a** Conceptual progression from the binary state selection and mechanically reconfigured sequential access to programmable spatiotemporal state distributions. **b** Space-time coding metasurface with two separately addressable polarization branches and four calibrated reflection settings per branch. The paired codes $C_x(\mathbf{R},t)$ and $C_y(\mathbf{R},t)$ synthesize the fields at $f_m$=$f_c$+$mf_s$, with the aperture coordinate $\mathbf{R}$, modulation time $t$, carrier frequency $f_c$ and modulation frequency $f_s$. **c** Temporal state trajectory with the fixed observation position and frequency. **d** Spatial state texture at a fixed preparation and frequency. **e** Spectral state distribution at a fixed preparation and observation position, generated by a shared code pair. The preparation index $\tau$ labels the successive complete code pairs. Sphere axes denote the normalized Stokes components.

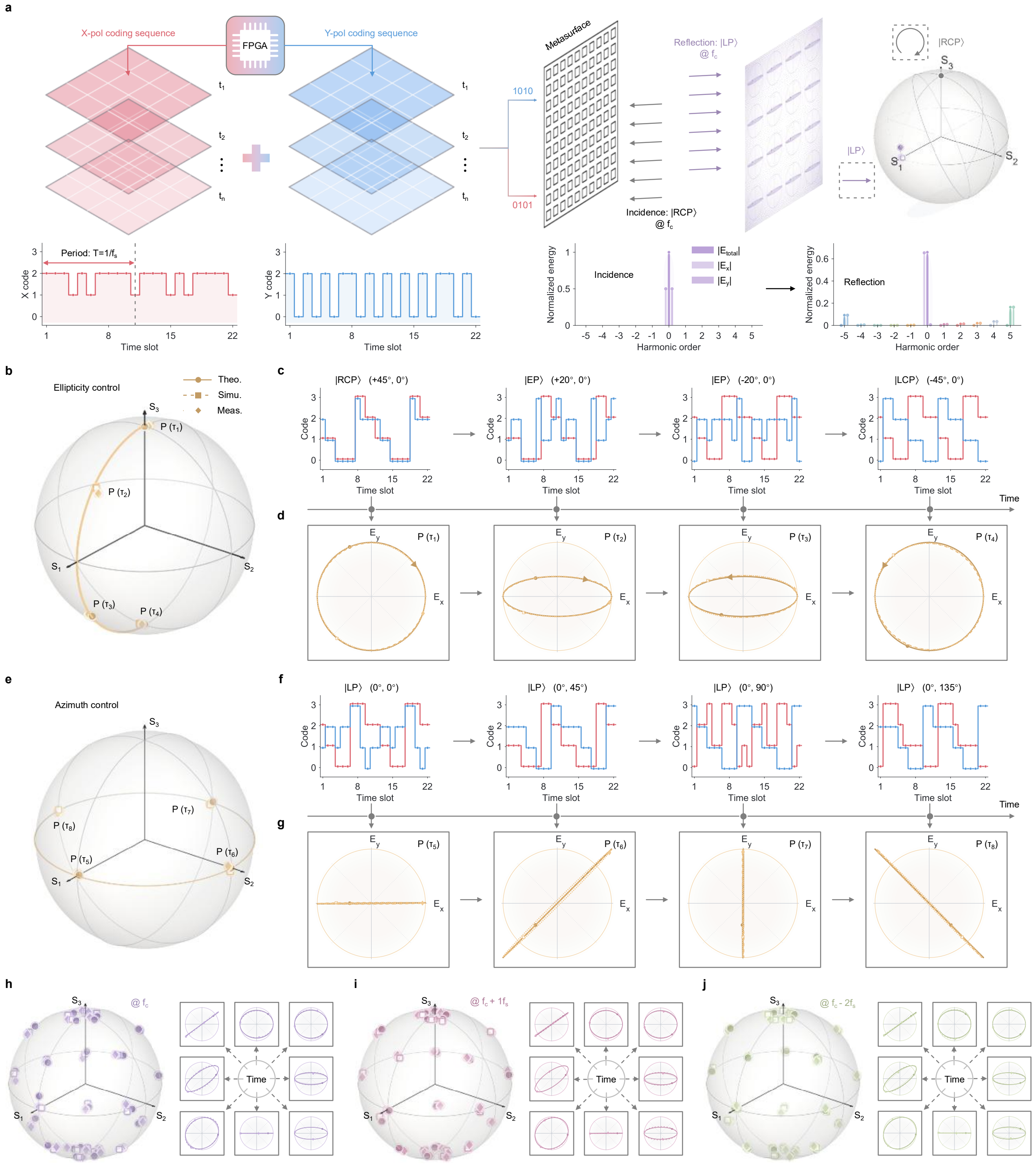

**Fig. 2 | Temporal interference and programmed paths on the state sphere.** **a** Paired temporal codes and the corresponding spectra for converting the circularly polarized incidence into a linearly polarized carrier component. Red and blue denote the $x$ and $y$ polarization branches. **b-d**, Meridional carrier trajectory (**b**), code pairs (**c**) and reconstructed ellipses (**d**) for the preparations $P(\tau_1)$ to $P(\tau_4)$. **e-g**, Equatorial trajectory (**e**), code pairs (**f**) and reconstructed ellipses (**g**) for $P(\tau_5)$ to $P(\tau_8)$. The angles $\chi_e$ and $\psi_e$ specify the ellipticity and azimuth, respectively, with the sphere latitude $2\chi_e$ and longitude $2\psi_e$; the azimuth is undefined at circular polarization. **h-j**, Separately optimized states at $f_c$ (**h**), $f_c$+$f_s$ (**i**) and $f_c$-2$f_s$ (**j**), with the representative ellipses and a distinct code pair for each target and order. Circles, squares and diamonds indicate theory, simulation and measurement, respectively. The slot coordinate resolves each period $T$=1/$f_s$, whereas $\tau$ indexes the successive preparations.

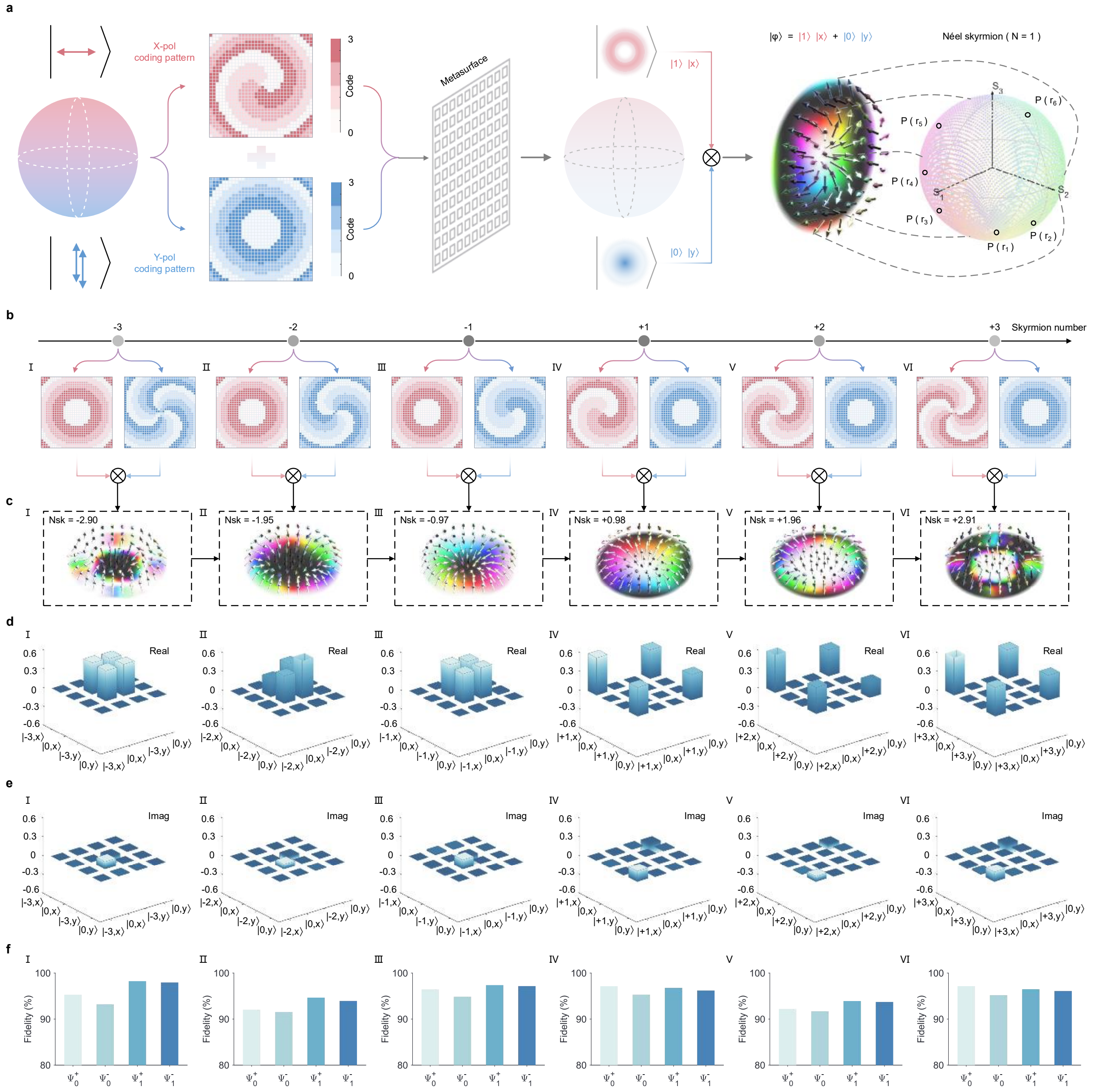

**Fig. 3 | Reconfigurable winding and modal coherence of spatial state fields.** **a** Spatial state synthesis from an $x$-polarized vortex component and a $y$-polarized non-vortex component, whose relative amplitude and phase define the local Poincaré coordinates. **b** Spatial code pairs for configurations I to VI with the target charges of $N_{\mathrm{sk}}^{\mathrm{tar}}$=-3,-2,-1,+1,+2,+3. Red and blue identify the $x$ and $y$ branches, with logical settings 0 to 3. **c** Reconstructed Stokes textures with finite-domain integrals -2.90,-1.95,-0.97,+0.98,+1.96,+2.91, respectively. **d,e** Real (**d**) and imaginary (**e**) parts of the normalized modal coherency matrices $\rho_{\mathrm{sub}}$ for representative preparations, with the axes identifying the spatial mode and polarization. **f** Conditional modal fidelities for four target superpositions prepared separately in each configuration. All matrices and target vectors use the same ordered basis within the selected spatial-mode and polarization subspace.

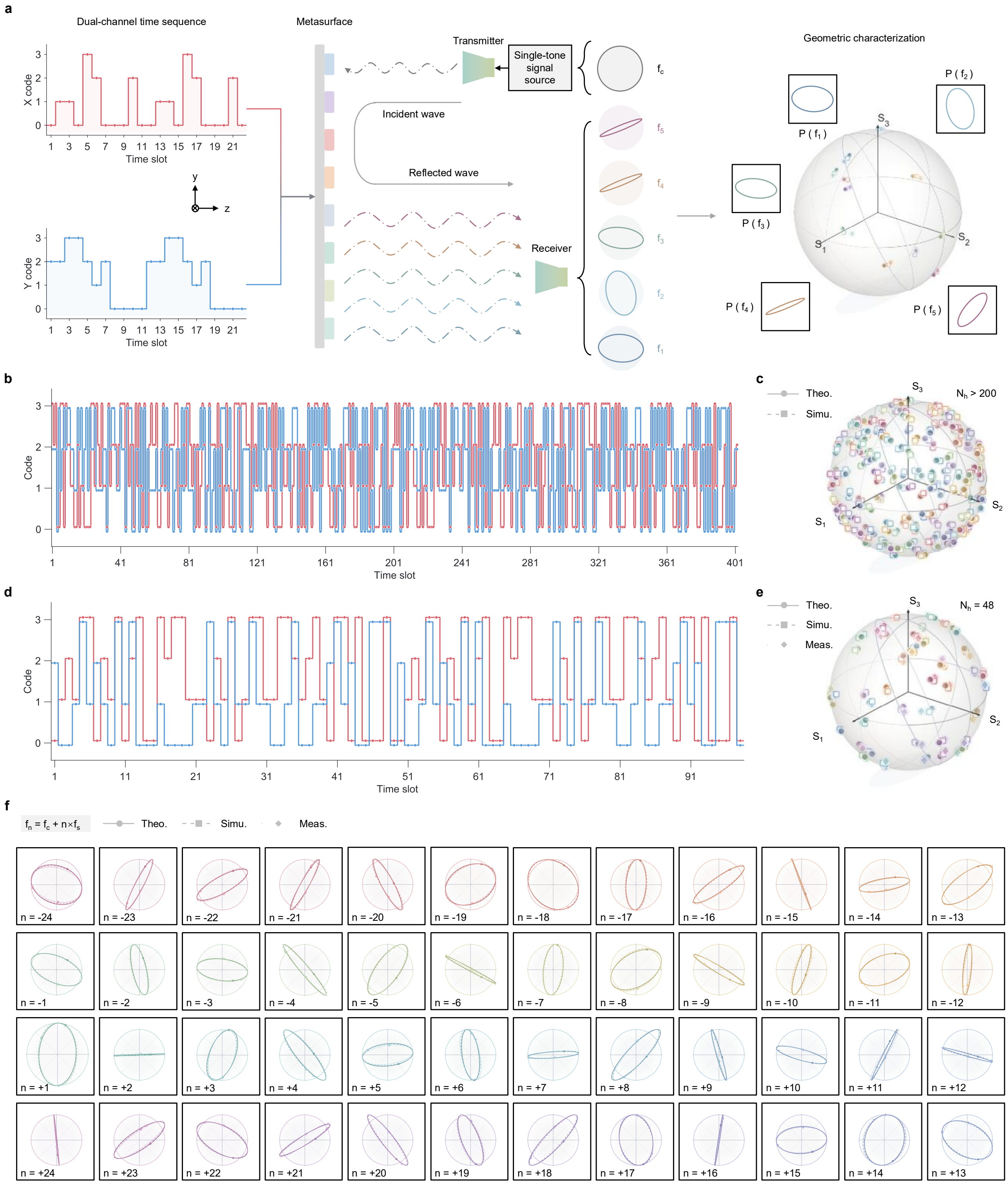

**Fig. 4 | Spectral state distributions from a shared temporal preparation. a** Paired periodic codes convert the incidence at $f_c$ into components at $f_m$=$f_c$+$mf_s$, whose normalized states are resolved at a common observation position and preparation. Red and blue denote the $x$ and $y$ code streams, with logical settings 0 to 3. **b,c** Numerical syntheses over more than 200 harmonic orders, showing the shared codes (**b**) and theoretical and simulated state coordinates (**c**). **d,e** Experimental syntheses over 48 nonzero orders, showing the shared codes (**d**) and theoretical, simulated and measured coordinates (**e**). Each distribution uses an unchanged code pair, and $N_h$ counts its displayed harmonic orders. **f** Normalized polarization ellipses for $m$=-24 to -1 and $m$=+1 to +24. Circles, squares and diamonds identify the theory, simulation and measurement, respectively. Ellipse orientation and axial ratio describe

the polarization geometry, with the handedness determined by the reconstructed complex fields; the normalization removes the absolute harmonic intensity.